\documentclass[usenatbib]{mn2e}
\usepackage{graphicx}
\usepackage{txfonts}
\usepackage{threeparttable}
\usepackage{color}
\usepackage{aas_macros}
\title[The sustainable growth of the first black holes]{The sustainable growth of the first black holes}
\author[Pezzulli et al.]{Edwige Pezzulli$^{1,2,3}$\thanks{E-mail:edwige.pezzulli@uniroma1.it}, Marta Volonteri$^{4}$, Raffaella Schneider$^{1,2}$, Rosa Valiante$^2$\\
$^{1}$Dipartimento di Fisica, Universit{\'a} di Roma  ``La Sapienza'', P.le Aldo Moro 2, 00185, Roma, Italy \\
$^{2}$INAF/Osservatorio Astronomico di Roma, Via di Frascati 33, 00040 Monte Porzio Catone, Italy\\
$^{3}$INFN, Sezione di Roma I, P.le Aldo Moro 2, 00185 Roma, Italy\\
${^4}$Institut d'Astrophysique de Paris, Sorbonne Universit{\'e}s, UPMC Univ. Paris 06 et CNRS, UMR 7095, F-75014, Paris, France}
\begin{document}

\date{20 June 2017}

\pagerange{\pageref{firstpage}--\pageref{lastpage}} \pubyear{2017}

\maketitle
\label{firstpage}

\begin{abstract}
Super-Eddington accretion has been suggested as a possible formation pathway of $10^9 \, M_\odot$ supermassive black holes (SMBHs) 800 Myr after the Big Bang. However, stellar feedback from BH seed progenitors and winds from BH accretion disks may decrease BH accretion rates.
In this work, we study the impact of these physical processes on the formation of $z \sim 6$ quasar, including new physical prescriptions in the cosmological, data-constrained semi-analytic model \textsc{GAMETE/QSOdust}. 
We find that the feedback produced by the first stellar progenitors on the surrounding does not play a relevant role in preventing SMBHs formation.
In order to grow the $z \gtrsim 6$ SMBHs, the accreted gas must efficiently lose angular momentum.
Moreover disk winds, easily originated in super-Eddington accretion regime, can strongly reduce duty cycles. This produces a decrease in the active fraction among the progenitors of $z\sim6$ bright quasars, reducing the probability to observe them.
\end{abstract}


\section{Introduction}

Observations of luminous ($L \gtrsim 10^{47} \rm erg/s$) quasars at $z \sim 6$ reveal that these objects host in their centres supermassive black hole (SMBH) with $M_{\rm BH} \gtrsim 10^{9} \, \rm M_{\odot}$. This poses strong constraints on theoretical models for the evolution of their less-massive progenitors (seeds). In fact, high-$z$ SMBHs must have formed in $\lesssim 1$ Gyr, which is the corresponding age of the Universe at those redshifts. How did the first black holes (BHs) seeds grow so fast is still an open question.

First BH seeds should have been born at $z \gtrsim 15$ and different physical mechanisms for their formation have been proposed. The first main scenario predicts \textit{light} seeds, consisting in Population~III (Pop~III) stellar remnants with mass $M_{\rm seed} \sim [10-1000] \, M_{\odot}$, formed at $z \gtrsim 20$ mostly in halos with $\rm T_{\rm vir} < 10^4 \, K$, called minihalos (\citealt{Abel2002,Bromm2002,Turk2009,Tanaka2009}). The second major channel predicts \textit{heavy} seeds of $10^5 - 10^6 \, M_\odot$ formed by the direct collapse of a protogalactic gas cloud in Lyman-$\alpha$ (Ly$\alpha$) cooling halos (i.e. halos with $\rm T_{\rm vir} \geq 10^4 \, K$) at $z \gtrsim 10$ (\citealt{Bromm2003,Begelman2006,Volonteri2006,Lodato2006}).
The birth-place of direct-collapse black holes (DCBHs) should be metal-free, to prevent metal-line cooling and fragmentation, and has to be illuminated by a strong Lyman Werner flux to efficiently photo-dissociate $\rm H_2$ molecules and prevent the gas from cooling and forming stars \citep{Omukai2008}.
In order to build up $z \sim 6$ SMBHs, DCBH scenario may represent a head start, which helps in explaining the existence of such massive, early objects, by starting from high-mass seeds. However, the physical conditions required to their formation seem to be rare (\citealt{Dijkstra2014, Habouzit2016, Chon2016, Valiante2016}, but see \citealt{Regan2017}).

On the other hand, forming high-$z$ quasars starting from \textit{light} seeds and assuming an Eddington limited growth would require uninterrupted gas accretion, which is quite unrealistic. 
In fact, feedback effects, produced by the accretion process itself, can strongly affect gas inflow in minihalos or, more generally, low-mass dark matter halos, resulting in negligible mass growth \citep{Johnson2007, Alvarez2009, Milosavljevic2009a, Madau2014}.
A possible solution is the occurrence of short, episodic super-Eddington accretion events (\citealt{Haiman2004, Shapiro2005, Volonteri2005, Pelupessy2007, Tanaka2009, Madau2014, Volonteri2015, Pezzulli2016}). Moreover, thanks to an early, efficient super-critical growth, it is possible to achieve in $\sim$ few Myr a BH mass comparable to what predicted by the direct collapse scenario ( \citealt{Madau2014, Lupi2015}).

In \citet[][, hereafter P16]{Pezzulli2016} it is shown that $\sim 80 \%$ of the mass of $z \sim 6$ SMBH with $M_{\rm BH} \sim 10^9 \, M_\odot$ is grown via super-critical accretion events, which represent the dominant contribution at $z \gtrsim 10$. In fact, such accretion regime is favoured in dense, gas-rich environments characterized by high column densities, which are common at high redshift.
On the contrary, the assumption of Eddington-limited accretion makes it impossible to reproduce the final SMBH mass.

This early super-Eddington accretion regime might provide an explanation for the current lack of faint AGN observations in the X-ray bands (\citealt{Treister2013, Weigel2015, Georgakakis2015, Cappelluti2016, Vito2016}). In fact, short episodes of mildly super-Eddington growth, followed by longer periods of quiescence may decrease the probability of observing BHs in active phases (\citealt{Pezzulli2017}, see also \citealt{Prieto2017}).

There are some physical processes that can suppress super-Eddington accretion in a cosmological context.
First of all, the rate at which seed BHs can grow, immediately following their formation, strongly depends on the feedback effects of their stellar progenitors. This may create gas poor environment surrounding the BH, giving rise to a delay on the early growth of the first seeds \citep{Johnson2007, Alvarez2009, Johnson2016}. 
Moreover, an important factor which limits the duration of super-Eddington accretion is the feedback produced by the accretion process on the disk itself.
In fact, a large fraction of the super-critical accretion power can drive disk winds, with a consequent loss of matter and, thus, a drop of the accretion rate \citep{Bisnovatyi-Kogan1977,Icke1980, Poutanen2007}.

In this work, we investigate the impact that the above mechanisms have on the early growth of the first BHs, assessing the feasibility of super-Eddington accretion as a channel for the formation of the first SMBHs.
To this aim, we study the relative impact of these hampering mechanisms for super-Eddington growth using the cosmological semi-analytic model presented in P16.

\section{Super-critical accretion flows}
\label{sec:2}

\begin{figure}
\centering
\includegraphics[width=8cm]{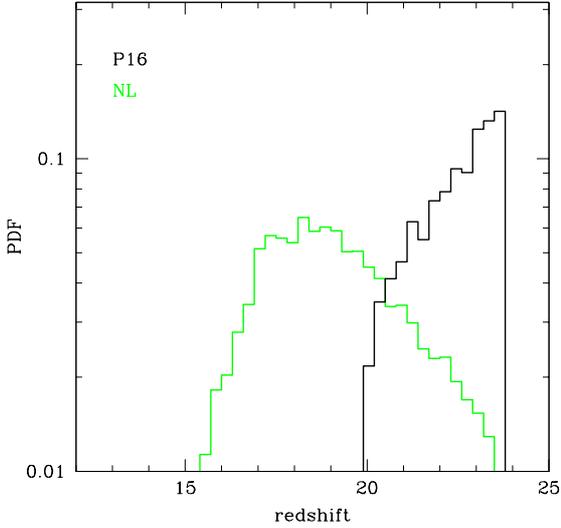}
\caption{Probability distribution function of $100 \, M_\odot$ BH seeds formation redshifts. PDF are averaged over 5 realizations. Green (black) histograms represent models with (NL) and without (P16) stellar feedback onto BH formation sites.}
\label{fig:z_distr}
\end{figure}

\begin{figure}
\centering
\includegraphics[width=8cm]{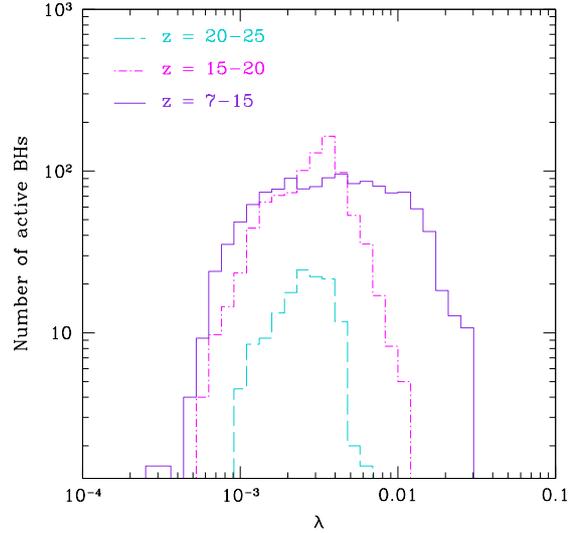}
\caption{Distribution of the parameter $\lambda$ in the redshift intervals $z = 20-25$ (turquoise, dashed), $z = 15-20$ (magenta, dashed-dotted), and $z = 7-15$ (violet, solid) for NL model.} 
\label{lambda_distr}
\end{figure}

\begin{figure*}
\includegraphics[width=16.8cm]{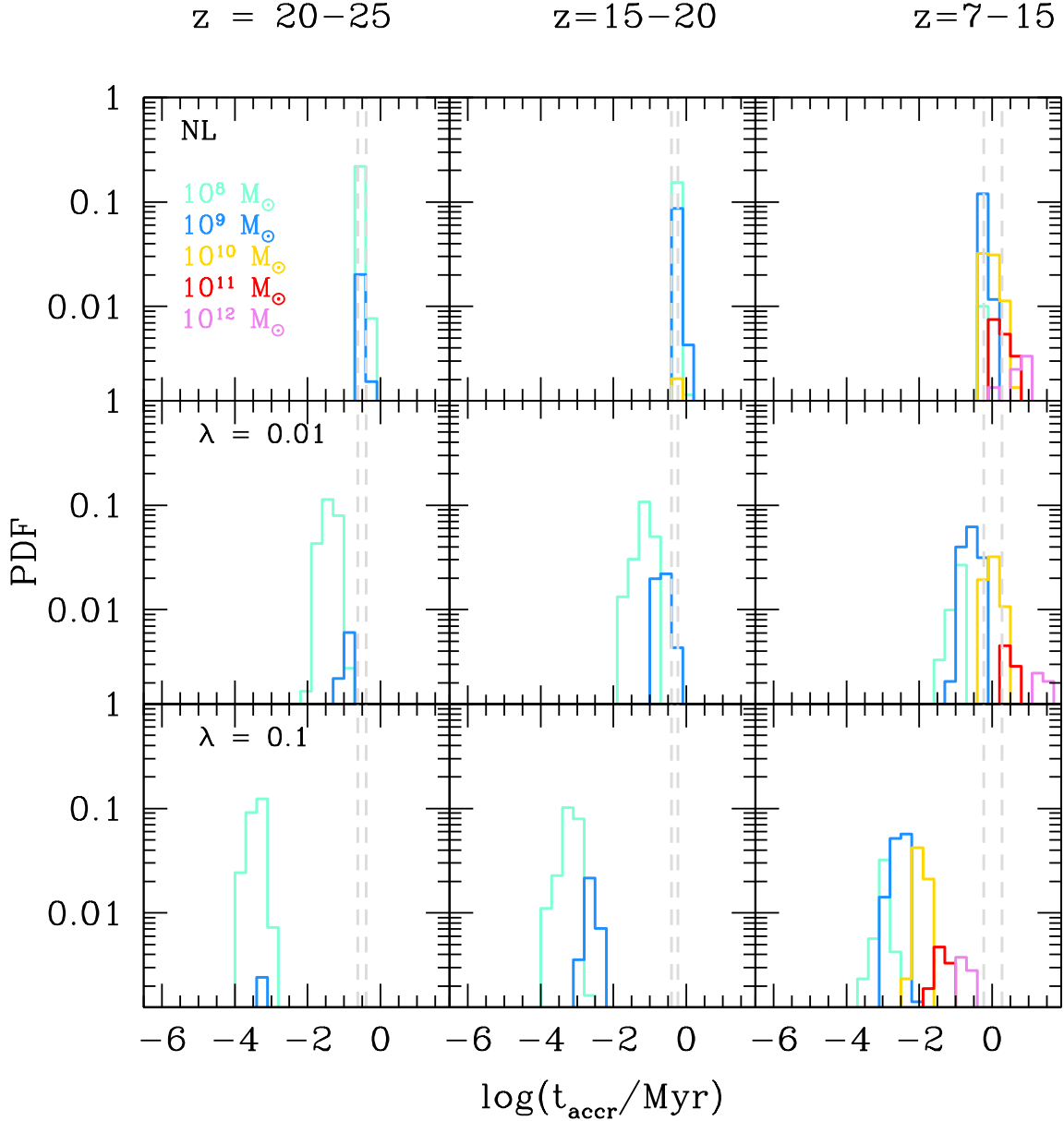}
\caption{Probability distribution function of the time duration of single super-Eddington accretion events for NL (top panels), L001 (middle panels) and L01 (bottom panels) models. Columns refer to different redshift intervals, $z = 20-25$ (left), $z = 15-20$ (center) and $z = 7-15$ (right), while colours indicate different mass of the BHs' DM host halos, as labelled in the top-left panel. Vertical dotted lines represent the maximum and minimum values of time resolution $\Delta t_{\rm r}$ of the simulation, in the related redshift interval.} 
\label{times}
\end{figure*}

\begin{figure*}
\centering
\includegraphics[width=8cm]{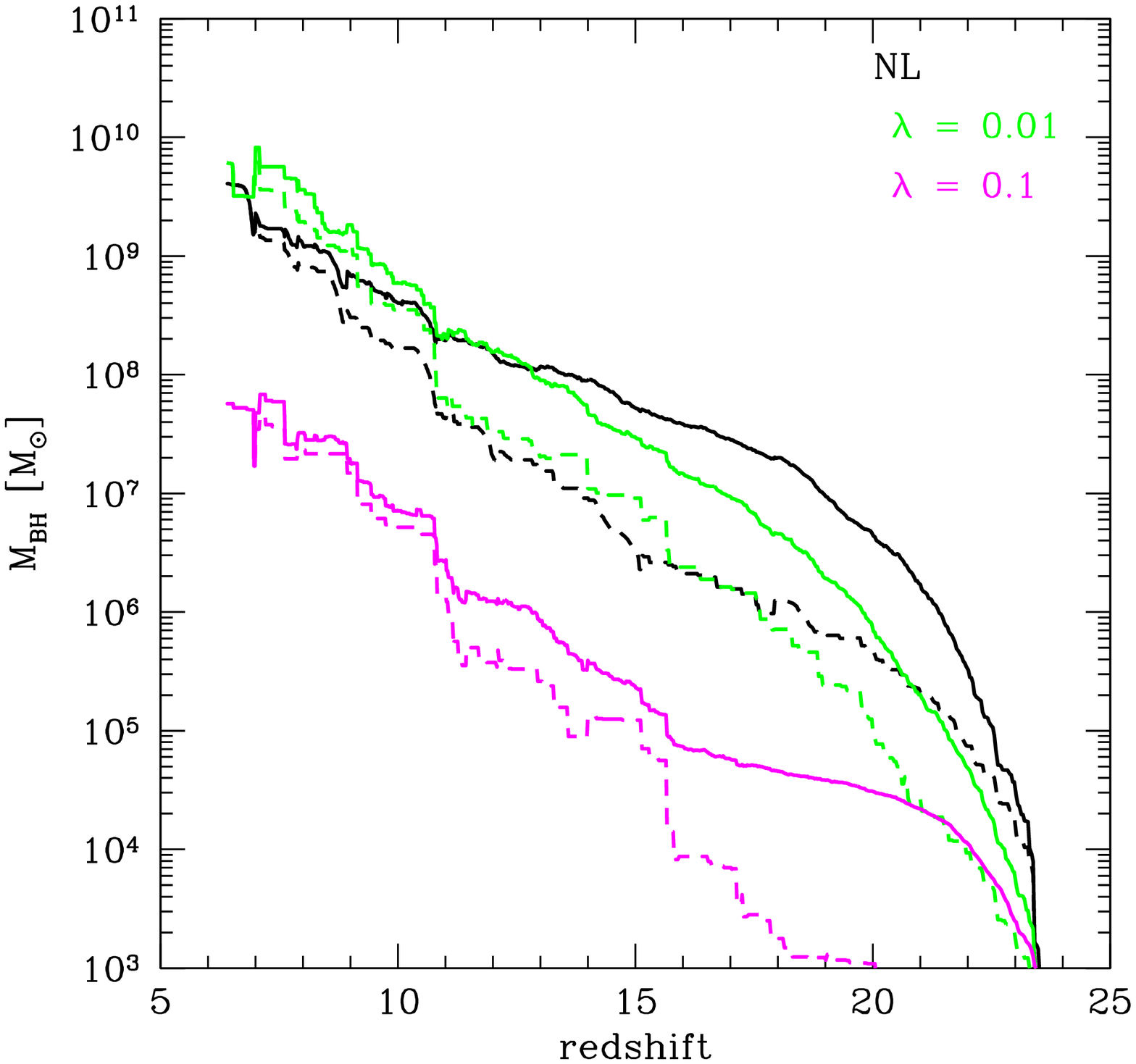}
\includegraphics[width=8cm]{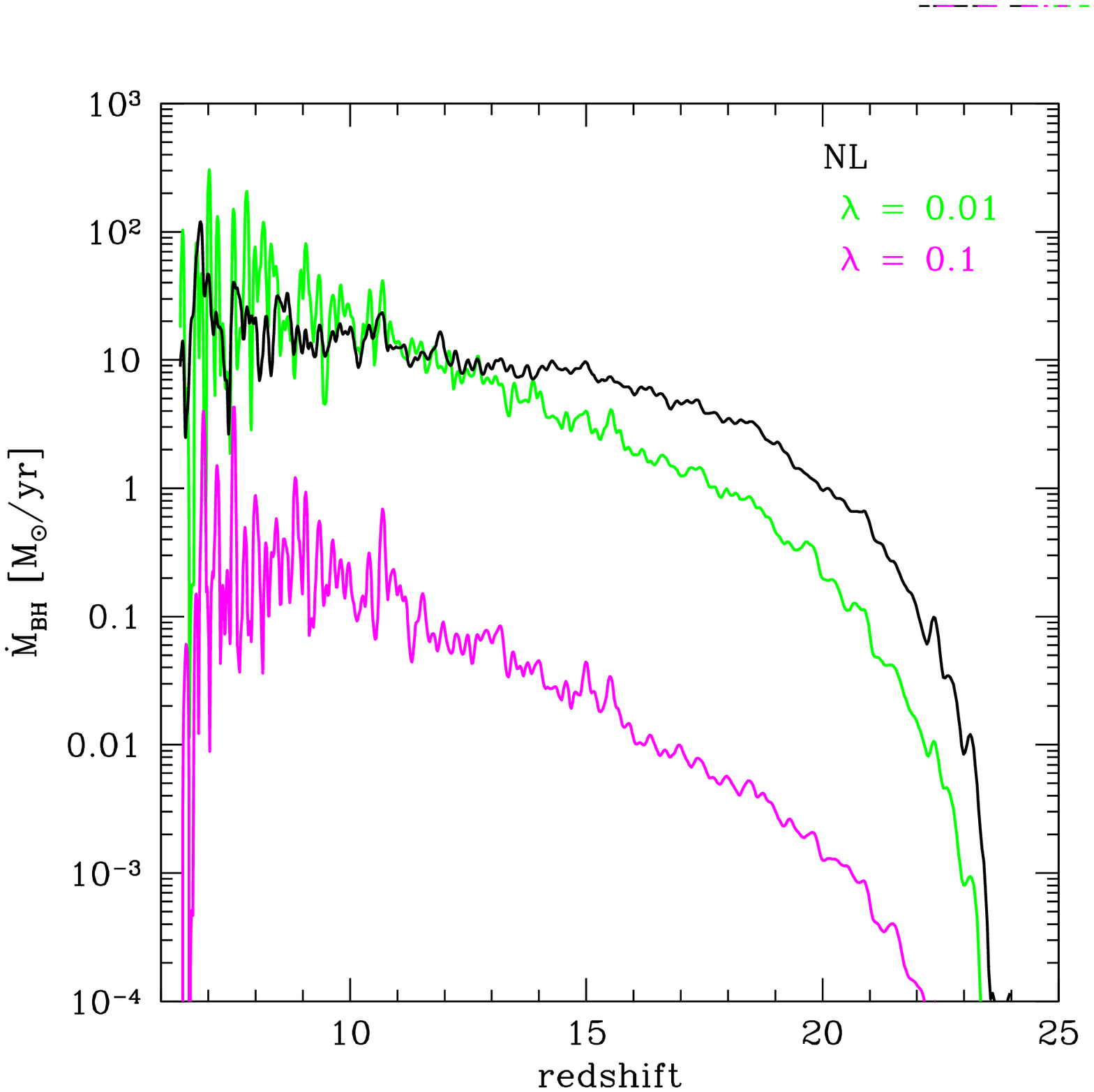}
\caption{Time evolution of the more massive (dashed lines) and total (solid lines) black hole mass (left panel) and black hole accretion rate (right panel) evolution for NL (black line), L001 (green line) and L01 (magenta line) models.} 
\label{bhevo}
\end{figure*} 

The model developed in P16 allows to reconstruct $N_r$ independent merger histories of a dark matter (DM) halo with $M_{\rm h} = 10^{13} M_\odot$,  assumed to host a typical $z \sim 6$ SMBH, like SDSS J1148 (e.g. \citealt{Fan2004}).

The time evolution of the mass of gas, stars, metals and dust in a two-phase interstellar medium (ISM) is self-consistently followed inside each progenitor galaxy and the model free parameters are fixed so as to reproduce some of the observed properties of the selected quasar (BH mass, gas mass, star formation rate, mass outflow rate radial profile). 

The hot diffuse gas, that we assume to fill each newly virialized DM halo, can gradually cool. 
For minihalos, we consider the contribution of $\rm H_2$, OI and CII cooling \citep{Valiante2016}, while for Ly$\alpha$-cooling halos the main cooling path is represented by atomic transitions. In quiescent evolution, the gas settles on a rotationally-supported disk.  It can be disrupted when a major merger ($M_{\rm h,1}/M_{\rm h,2} = \mu \geq 1/4$) occurs, forming a bulge structure, for which we adopt an Hernquist profile \citep{Hernquist1990}. 

In the model introduced in P16, we assume BH seeds to form with a constant mass of $100 \, M_\odot$ as remnants of Pop~III stars in halos with $Z \leq Z_{\rm cr} = 10^{-4} \, Z_\odot$ \citep{Valiante2016}, without considering any stellar radiative feedback effect produced by the first luminous BH progenitors on their environment. 

The BH can grow through gas accretion from the surrounding medium and via mergers with other BHs. 
Our prescription allows to consider quiescent and enhanced accretion, following merger-driven infall of cold gas, which loses angular momentum due to torque interactions between galaxies.
We model the accretion rate to be proportional to the cold gas mass in the bulge $M_{\rm b}$, and inversely proportional to the bulge dynamical time-scale $\tau_{\rm b}$:

\begin{equation}
\dot{M}_{\rm accr} = \frac{f_{\rm accr} M_{\rm b}}{\tau_{\rm b}},
\end{equation}

\noindent where $f_{\rm accr} = \beta f(\mu)$, with $\beta = 0.03$ in the reference model and $f(\mu) = \max[1, 1+2.5(\mu - 0.1)]$, so that mergers with $\mu \leq 0.1$ do not trigger bursts of gas accretion. 

At high accretion rates, the standard \textit{thin} disk model is no longer valid.
Therefore, the bolometric luminosity $L_{\rm bol}$ produced by the accretion process has been computed starting from the numerical solution of the relativistic slim accretion disk obtained by \citet{Sadowski2009}, adopting the fit presented in \citet{Madau2014}. This model predicts mildly super-Eddington luminosities even when the accretion rate is highly super-critical, limiting the impact of the feedback onto the host galaxy.
The energy released by the AGN can then couple with the ISM. We consider energy-driven feedback, which produces powerful galactic-scale outflows, and SN-driven winds, computing the SN rate explosion for each galaxy according to the formation rate, age and initial mass function of its stellar population \citep{deBennassuti2014,Valiante2014}. 

Finally, in BH merging events, the newly formed BH can receive a large center-of-mass recoil due to the net linear momentum carried by the asymmetric gravitational wave \citep{Campanelli2007,Baker2008}. We take into account this effect, computing the \textit{kick} velocities following \citet{Tanaka2009}, under the assumption of a random distribution of BH spins and angles between the BH spin and the binary orbital angular momentum vectors.

We refer the reader to P16 for a more detailed description of the model.
In the following paragraphs, we discuss the new features introduced in the model, i.e. the inclusion of the first stellar BH progenitors feedback on the surrounding gas, and a time-scale for the duration of a super-Eddington accretion event.

\subsection{Seeding prescription}


For each newly formed galaxy, we compute the star formation rate in the disk and in the bulge as $\dot{M}^\star_{\rm d,b} \propto M_{\rm d,b}/\tau_{\rm d,b}$, where $M_{\rm d,b}$ and $\tau_{\rm d,b}$ are the gas mass and the dynamical time of the disk (labelled 'd') and bulge ('b'), respectively (see section 2.2.1 in P16 for further details).

Following \citet{Valiante2016}, we assume Pop~III stars to form when $Z < Z_{\rm cr} = 10^{-4} \, Z_{\odot}$ in the mass range $[10 - 300]\, M_\odot$ according to a Larson IMF \citep{Larson1998}:

\begin{equation}
\Phi(m_\star) = \frac{dN(m_\star)}{dm_\star} \propto m_\star^{\alpha - 1} e^{-m_\star/m_{ch}}, 
\end{equation}

\noindent with $\alpha = -1.35$, $m_{ch} = 20 \, M_\odot$ (\citealt{deBennassuti2014, Valiante2016}).

For non-rotating stars with $Z = 0$, a $M_{\rm seed} \sim 100 \, M_{\odot}$ BH is expected to form from $M_\star \gtrsim 260 \, M_\odot$ \citep{Valiante2016}.
We do not consider as light seeds BHs forming from $[40-140] \, M_\odot$ progenitors because lighter BHs are not expected to settle steadily in the minimum of the potential well, due to stellar interactions (\citealt{Volonteri2010}). Moreover, we do not take into account stars with masses of $M_\star = [140-260] \, M_\odot$ , that are expected to explode as pair instability supernovae, leaving no remnants \citep{Heger2003,Takahashi2016}. 

The probability to find a BH seed with, \textit{at least},  $\sim 100 \, M_\odot$, after a single star formation episode is,
\begin{equation}
f_{\rm seed} = \frac{\int_{260}^{300} m_\star \Phi(m_\star) \, dm_\star}{\int_{10}^{300} m_\star \Phi(m_\star) \, dm_\star}.
\end{equation}

\noindent Based on results obtained by \citet{Valiante2016} through random sampling of the IMF, the condition  $f_{\rm seed} \sim 1$ requires a minimum stellar mass formed in a single burst of $1000 \, M_\odot$. Thus, conservatively, we assume that one $100 \, M_\odot$ BH seed forms after a star-formation episode only if the total stellar mass formed $\Delta M_\star$ is $\geq 10^3 \, M_\odot$.

\subsection{Stellar progenitors feedback}

The stellar progenitors of the first BHs are massive primordial stars, expected to form in minihalos. 
Their large luminosities, with a huge production of ionizing radiation for few Myr before their collapse (e.g. \citealt{Schaerer2002}), can couple with the surrounding gas and heat it above the virial temperature of the host dark matter halo. As a result, BH seeds likely form in low-density $\rm HII$ region (e.g. \citealt{Whalen2004, Alvarez2006}), with consequent low gas accretion rates (\citealt{Alvarez2009, Johnson2013,Johnson2016}). Due to this radiative feedback in minihalos, the newborn BH may wait up to $100 \, \rm Myr$ before starting to accrete efficiently.

Another important impact on the early BH growth is produced by SN explosions of massive primordial stars, which can provide a strong limit to the gas reservoir from which Pop~III relic BHs can accrete.

To take into account these negative feedback effects, we assume that, following each Pop~III star formation burst, all the gas is blown out of the galaxy, in the intergalactic medium (IGM).
In addition, to mimic the impact of photo-ionization and heating, which affect the large-scale inflow, we assume that gas accretion from the IGM is inhibited as long as the virial temperature of the host halo remains $T_{\rm vir} < 10^4 \, \rm K$. 
Furthermore, feedback produced by the first stars is strong enough to prevent further cooling and star formation within its host minihalo for the subsequent $200 \, \rm Myr$ (\citealt{Alvarez2009}). For this reason, we suppress gas cooling in minihalos after the first star formation event, and relax this constraint only for halos with virial temperature $T_{\rm vir} \geq 10^4 \, \rm K$.

\subsection{The duration of super-Eddington accretion events}

Idealistic slim accretion disk model predicts that a large fraction of the radiation produced by the accretion process can be advected into the BH instead of escaping. In fact, it is possible to define a radius $R_{\rm pt}$ within which the trapping of radiation becomes relevant.
Trapping of radiation occurs in regions of the accretion disk for which the diffuse time scales $t_{\rm diff}(r)$ is larger than the accretion time $t_{\rm accr}(r)$. Imposing $t_{\rm diff} = t_{\rm accr}$ it is possible define the photon trapping radius $R_{\rm pt}$ \citep{Ohsuga2002} :

\begin{equation}
R_{\rm pt} = \frac{3}{2}\dot{m}\, h  \, R_{s},
\end{equation}

\noindent where $R_{s} = 2GM_{\rm BH}/c^2$ is the Schwarzschild radius, $\dot{m} = \dot{M}
_{\rm accr}/\dot{M}_{\rm Edd}$ is the Eddington accretion ratio and $h=H/r$ is the ratio between the half disk-thickness $H$ and the disk radius $r$. Since $h \approx 1$ in radiation pressure dominated regions, we assume $h = 2/3$ so that $R_{\rm pt} = R_{s}\dot{m}$.

In realistic cases, however, the accretion process can be suppressed. The outward angular momentum transport, necessary for accretion, also involves a transport of energy. This produces unbounding of gas far from the BH, thus less gas has the possibility to reach it.
Moreover, a significant fraction of the accretion power in super-critical flows may drive disk winds. In fact, at large luminosities, flows are supported by radiation pressure, which is likely to induce outflows \citep{Shakura1973, Bisnovatyi-Kogan1977,Icke1980, Ohsuga2005, Poutanen2007}.
Results of recent simulations suggest that the mass lost due to disk winds becomes relevant only as photon trapping becomes less important, i.e. in the outer region of the disk \citep{Ohsuga2007,  Takeuchi2009, Begelman2012,Sadowski2014}.
As already discussed in \citet{Volonteri2015}, it is thus possible to assume that a significant disk wind is produced only after the disk radius has reached some significant fraction of the trapping radius.
When this occurs, the mass lost to the outflow reduces the gas accretion rate, which can drop to $10-20 \%$ of the inflow rate (e.g. \citealt{Ohsuga2007}), decelerating the BH growth.
In addition, the mass outflow increases with the disk radius (\citealt{Volonteri2015}), so that both effects can eventually quench black hole growth once the trapping radius is reached (see also \citealt{Volonteri2005, Volonteri2015}).

Following \citet{Volonteri2015}, we assume that once the disk radius $R_{\rm d}$ reaches $R_{\rm pt}$, the disk is blown away, and the accretion process is no longer sustained.
This reflects into a condition on the maximum time for which super-Eddington accretion can be sustained\footnote{Being the disk radius $R_{\rm d} = \lambda^2 R_{\rm g} = \lambda^2 G M_{\rm BH}/\sigma^2$, and the Eddington luminosity $L_{\rm Edd} = t_{\rm Edd}/(M_{\rm BH} c^2)$, approximating $M_{\rm BH} = \dot{M}_{\rm BH} t$, the condition $R_{\rm d} \leq R_{\rm pt}$ turns into the inequality $(\lambda c/\sigma)^2 (M_{\rm BH}/2t_{\rm Edd} \dot{M}_{\rm BH}) \leq 1$.} \citep{Volonteri2015}:

\begin{equation}
t_{\rm accr} = 2 \lambda^{-2} \left( \frac{\sigma}{c} \right)^2 t_{\rm Edd},
\label{Eqtime}
\end{equation}

\noindent where $t_{\rm Edd} = 0.45$ Gyr is the Eddington time, $\lambda \leq 1$ is the fraction of angular momentum retained by the gas and $\sigma$ is the gas velocity dispersion.
The parameter $\lambda$ is defined as the specific angular momentum $\ell_g$ of matter crossing the BH sphere of influence, normalized to the Keplerian value, i.e. $\lambda = \ell_g/\sqrt{GM_{\rm BH}R_{\rm g}}$, where $R_{\rm g} = GM_{\rm BH}/\sigma^2$.

Since $R_{\rm d} \propto \lambda^2$, smaller values of $\lambda$ lead to smaller disk sizes and hence to a prolonged phase of super-Eddington accretion, $t_{\rm accr}$. 


For the present study we investigate two different values, $\lambda = 0.01$ and $\lambda = 0.1$. The latter is suggested by studies of angular momentum losses for gas feeding SMBHs during galaxy mergers. 
\citet{Capelo2015} find $\lambda < 0.5$ (with mean and median values of 0.28 and 0.27, respectively), in simulations with gas softening length of 20 pc. The former represent a more optimistic, but not extreme, case \citep[see][for a discussion]{Begelman2017}.

\section{Results}

In this section, we explore the impact of stellar feedback and of the disk outflow comparing the results of
the new models with those found in P16 where the above effects were not considered. Models with stellar feedback and $\lambda = 0.1$ and 0.01 have been labelled as L01 and L001, respectively. The model P16 described in Sec. \ref{sec:2}, including stellar feedback and no disk outflow has been labelled NL. This implies that the only difference between L01 (or L001) and NL resides in accounting or not for disk winds effects.
For each model, the results must be intended as averaged over $N_r = 5$ simulations.

\subsection{The impact of Stellar feedback}

Figure \ref{fig:z_distr} shows the redshift distribution of newly formed BH seeds with (green histograms, NL model) and without (black histograms, P16 model) the effect of stellar feedback. In the no-feedback case, due to efficient metal enrichment, Pop~III star formation becomes negligible below $z \sim 20$. The inclusion of stellar feedback causes a shift of BH seed formation to lower redshift.
Moreover, while in the no-feedback model we find $\sim 90 \%$ of BH-seeds hosts are minihalos, once feedback is considered native galaxies are mostly Ly$\alpha$-cooling halos. 
This stems from the condition that a $100 \, M_\odot$ BH remnant requires a minimum Pop~III stellar mass of $\Delta M_{\star} \sim 10^3 \, M_\odot$ formed in a single burst, which can be hardly accomplished in minihalos, due to the low-efficiency feedback-limited star formation. 
The effect is that Pop~III stars sterilize minihalos, without giving birth to a BH seed \citep{Ferrara2014}.
Once minihalos have grown enough mass to exceed $T_{\rm vir} = 10^4 \, \rm K$, gas cooling is more efficient and $100 \, M_{\odot}$ BH seeds have a larger probability to form.
As a result, BH seeds continue to form down to $z \sim 15$ in the NL model, in good agreement with what found in \citet{Valiante2016}. 

\subsection{Super-Eddington duration}

To understand the impact of the duration of super-Eddington accretion episodes on high-$z$ SMBHs growth, we have compared the L01 and L001 cases with the NL model. In the NL model, disk winds effects are not considered. Thus, the accreting event - and its lifetime - depends only on the presence, in a galaxy, of a BH surrounded by a gas reservoir. 
Since there is no apriori constraint on the accretion time-scale, it is possible to invert Equation \ref{Eqtime} and obtain the distribution of $\lambda$ values shown in Figure \ref{lambda_distr}.

Model NL results in values of $\lambda$ smaller than assumed in models L01 and L001, with $ 10^{-4} \lesssim \lambda \lesssim 10^{-1}$. We find slightly increasing values of $\lambda$ for decreasing redshift, with wider distributions at lower $z$. This effect is dominated by an increasing dispersion in the values of $\sigma$ for decreasing redshift.
In fact, the duration of super-Eddington accretion, $t_{\rm accr}$, follows a narrow distribution around the time resolution $\rm \Delta t_{\rm r}$ of the simulation at the corresponding redshift, with BHs accreting at most $\sim$ few times $\Delta t_{\rm r}$ (see the top row of Fig. \ref{times}). These short durations are consequence of the rapid depletion of gas produced by efficient super-Eddington accretion, which represents the dominant contribution at all but the latest redshift of the SMBH evolution (see P16 for details).
Conversely, in models L001 and L01 we have limited super-Eddington accretion to $\rm t_{\rm accr}$ as obtained from Equation \ref{Eqtime}, with resulting distributions shown in the middle (L001) and bottom (L01) panels of Figure \ref{times}.
It is interesting to note that, under the assumption of $\lambda = 0.01$ or $\lambda = 0.1$, the accretion time-scales at $z > 15$ are shorter than adopted in P16 (hence in the NL model). 
In fact, larger values of $\lambda$ implies less compact objects and, thus, larger values of $R_{\rm d}$. This gives rise to shorter super-Eddington accretion episodes.
For $z = 20 - 25$, where the entire population of active BHs is accreting at super-critical regimes, the L01 model predicts an accretion-time distribution peaking around $\rm t_{\rm accr} \sim 100 \, yr$, to be compared with $\rm t_{\rm accr} \sim 0.01$ ($\sim 1$) $\rm Myr$ in L001 (NL) model, respectively. 
For lower $z$, the contribution of active galaxies with large gas velocity dispersion $\sigma$ becomes relevant, and the accretion times $\rm t_{\rm accr}$ become larger. For instance, in the L001 model it is possible to find BHs accreting for longer times (up to $\sim 30 \rm \, Myr$) with respect to the NL model, where $\rm t_{\rm accr} \sim 1 \, Myr$.

The distribution of $t_{\rm accr}$ shows an increasing trend with increasing dark matter halo mass. This effect is negligible in the narrow distribution predicted by model NL. In models L01 and L001, instead, one order of magnitude increase in dark matter halo masses corresponds to increasing $\gtrsim$ half order of magnitude accretion time-scales $t_{\rm accr}$.

It is interesting to compare how different assumptions on $\lambda$ affect the BH mass growth.
In the left panel of Figure \ref{bhevo} we show the evolution of the total (solid) BH mass, summing over all the progenitors present in the simulation at a given redshift. Dashed lines represent the time evolution of the most massive BH that powers the $z \sim 6$ quasar.
At high-$z$, the difference in the total BH mass between NL and L001 models is about one order of magnitude, as a consequence of different total black hole accretion rates (Hanning smoothed), shown in the right panel of Figure \ref{bhevo}. 
This quantity is computed as $\dot{M}_{\rm BH} = \Delta M_{\rm BH}/\Delta t_{\rm r}$, i.e. as the average BH mass increase in the simulation time-step $\Delta t_{\rm r}$, even if $t_{\rm accr} < \Delta t_{\rm r}$. Hence, lower BH accretion rates are a consequence of the lower $t_{\rm accr}$. More gas is retained by dark matter halos due to reduced AGN feedback effects, leading to larger BH accretion rates at later times. As a results, in model L001 the total BH mass follows a steeper evolution at $z<10$ compared to model NL, reaching a factor 2 larger value at $z=6.4$.
 
Conversely, the accretion time-scales, $t_{\rm accr}$, in the L01 model are too small to allow an efficient BH mass growth. 
Almost all the BHs present in model L01 accrete at super-Eddington rates for $t_{\rm accr} \sim 100 - 1000 \, \rm yr$. This leads to a BH mass growth from $\sim 10^5 \, M_\odot$ to $10^6 \, M_\odot$ between $z = 15-22$ and to a final BH mass $\sim$ 2 orders of magnitude lower than predicted by L001 and NL models.

\section{Conclusions}

Many models invoke super-Eddington accretion onto the first black holes as a possible route to form high-$z$ SMBHs \citep{Volonteri2005, Wyithe2012, Madau2014, Alexander2014, Volonteri2015, Inayoshi2015, Sakurai2016, Ryu2016, Begelman2017}. In P16, we have shown that super-Eddington accretion is required to form a $\sim 10^9 \, M_\odot$ SMBH at $z \sim 6$ starting from $\sim 100 \, M_\odot$ BH remnants of very massive Pop~III stars.
However, there are different mechanisms which can suppress early super-critical accretion. Feedback effects from the stellar progenitors can strongly affect the gas density around the newborn black holes, reducing the efficiency of gas accretion.
In addition, the onset of disk winds can suppress BH growth, setting a maximum time-scale for sustainable super-Eddington accretion. 

In this work, we used the cosmological, data-constrained semi-analytic model \textsc{GAMETE/QSOdust}, described in P16, to estimate the impact of these two physical processes on SMBHs formation at $z > 6$. 

We find that the influence of stellar feedback on the surroundings produce a delay on BH seeds formation, shifting their redshift distribution from $z \gtrsim 20$ to $z \gtrsim 15$. However, despite the very conservative assumptions made to maximize stellar feedback effects, we find that this delay does not prevent neither the growth of high-z SMBHs, nor the possibility of their BH progenitors to accrete at super-Eddington rates.

The impact of disk outflows, and the associated reduction of the duration of super-Eddington accretion episodes, strongly depends on the angular momentum of gas joining the accretion disk. 
Assuming that disk winds suppress BH accretion when the disk radius becomes comparable to the photon trapping radius, the result relies on the value of $\lambda$, which represents the fraction of angular momentum retained by the gas.
For $\lambda = 0.1$, $t_{\rm accr} \sim 100 - 10^4$ yr at $z>15$, too short to allow the SMBH to grow efficiently, and at $z \sim 6$ the final SMBH mass is $\sim$ 2 orders of magnitude lower than what obtained in the model where disk winds are neglected.
For $\lambda = 0.01$, instead, super-critical accretion events are sustained for time-scales $\sim 10^4- 10^6$ yr.
This suppresses the early growth phase, but the larger gas mass retained allows a steeper growth of the SMBH mass at later times.

The implication of this study is that the accreted gas must efficiently loose angular momentum to enable super-Eddington growth of the first SMBHs from light BH seeds. If $\lambda < 0.01$, super-Eddington accretion has a very short duty cycle, with $t_{\rm accr} \ll$ Myr at $z>15$ and for $\sim 0.1$ Myr for $z = 7-15$. This decreases the active fraction of high-$z$ BHs and further strengthens the conclusions of \citet{Pezzulli2017}, that the higher-redshift progenitors of $z\sim6$ quasars are difficult to observe "in the act", as the short and intermittent super-critical accretion events imply a low fraction of active black holes.

\section*{Acknowledgments}
We are grateful to the Referee, John Regan, for his useful suggestions and comments.

EP acknowledges the kind hospitality of the IAP, where
part of this work has been developed.
The research leading to these results has received funding from the European Research Council under the European Union’s Seventh Framework Programme (FP/2007-2013)~/~ERC Grant Agreement n. 306476.









\bibliography{bibliography}
\label{lastpage}

\end{document}